\documentclass[12pt]{iopart}

\usepackage{graphicx}

\newcommand{\BQ}{\begin{equation}}
\newcommand{\EQ}{\end{equation}}
\newcommand{\BQA}{\begin{eqnarray}}
\newcommand{\EQA}{\end{eqnarray}}
\newcommand{\be}{\begin{eqnarray}}
\newcommand{\ee}{\end{eqnarray}}

\newcommand{\kk}{k_\perp}

\newcommand{\bm}[1]{{\bf #1}}

\def\simge{\mathrel{%
   \rlap{\raise 0.511ex \hbox{$>$}}{\lower 0.511ex \hbox{$\sim$}}}}
\def\simle{\mathrel{
   \rlap{\raise 0.511ex \hbox{$<$}}{\lower 0.511ex \hbox{$\sim$}}}}

\begin{document}
\begin{flushright}
SPhT-T04/034
\end{flushright}

\vspace*{-6mm}
\title[CGC and BFKL in DIS at small $x$]
{Color Glass Condensate and BFKL dynamics\\ in deep inelastic scattering 
at small $x$}
\vspace*{-2mm}
\author{Kazunori Itakura}
\vspace*{-2mm}
\address{Service de Physique Th\'eorique, CEA Saclay,
        F-91191 Gif-sur-Yvette, France}

\begin{abstract}
The proton structure function $F_2(x,Q^2)$ 
for $x\le 10^{-2}$ and $0.045\le Q^2 \le 45\,{\rm GeV}^2$, 
measured in the deep inelastic scattering at HERA, 
can be well described within the framework of the Color Glass Condensate.
\end{abstract}
\vspace*{-0.7cm}
\section{Introduction}
\vspace*{-3mm}
There has been a surge of theoretical and experimental interest 
in the ``Color Glass Condensate (CGC)" which appears in 
the new perturbative regime of QCD relevant for high energy 
scattering \cite{Review}. This new state is characterized by 
 high density gluons whose longitudinal momenta are very 
small compared to the total momentum of the parent hadron 
 (such gluons are called ``small-$x$" gluons since 
the ratio of the momenta is denoted as $x$). 
The gluon density is typically as large as  ${\cal O}(1/\alpha_s)$, 
and cannot be too large (``saturated") so that the 
unitarity of physical cross sections is ensured. 

Recent rapid 
theoretical progress in understanding the physics of CGC is 
triggered by the experiments currently investigated at HERA 
(DESY) and RHIC (BNL). The relevant process at HERA is the deep 
inelastic scattering (DIS) of an electron
off a proton, while at RHIC it is more complicated Au-Au or 
deuteron-Au collision. 
These two apparently different experiments 
(different in complexity and energy) are nevertheless closely 
related to each other in the context of CGC through the 
``{\it universality}" of the hadron/nucleus wavefunctions or the 
gluon distributions. 
Here we mean differently by the word ``universality'' than in 
the usual sense. Namely, in the saturated regime at 
high energy, the gluon distributions of the proton and nucleus are 
both described by the 
{\it same function} of the ratio of the transverse momentum 
of gluons $\kk$ to the {\it saturation momentum} $Q_s$, 
which is the (inverse of) typical transverse size of gluons 
(for more explanations,
see for example, Refs.~\cite{IIT,SAT}). 
Therefore, one will be able to 'translate' the HERA physics 
for protons into the RHIC physics for gold nuclei. 
Furthermore, the saturation scales in two experiments 
are accidentally of the same order because of its particular dependencies 
upon $x$ and the atomic number $A$, i.e., 
$Q_s^2(A,x)\propto A^{1/3}(1/x)^\lambda\sim (A/x)^{0.3}$.
Indeed, at HERA, $A=1$ and $x\sim 10^{-4}$ while at RHIC, $A\simeq 200$ 
and typically $x\sim 10^{-2}$. This fact also strengthens 
the importance of understanding the HERA physics in relation 
to the RHIC physics. In this talk, I will show that the recent HERA data 
\cite{New_HERA} at small-$x$ with not so large $Q^2$ is consistent 
with the current picture of the CGC (for more detail, see Ref.~\cite{IIM}), 
which, according to the discussion above, 
suggests the importance of CGC at RHIC.

\section{DIS at small $x$ and previous fits with saturation models}
\vspace*{-3mm}
DIS at small-$x$ looks simple in the ``dipole picture" which leads to
an intuitively understandable factorization formula. The scattering 
between the virtual photon $\gamma^*$ (emitted from the projectile 
electron) and the proton is seen as the dissociation of $\gamma^*$ 
into a quark-antiquark pair (the ``color dipole'') followed by the 
interaction of this dipole with the color fields in the 
proton. Then one can write the $F_2$ structure function as
\be\label{sigmagamma}
F_2(x,Q^2)= \frac{Q^2}{4 \pi^2 \alpha_{\rm em}}\,
\sum_{T,L}\int dz\,d^2{\bm r}\,|\Psi_{T,L}(z,{\bm r},Q^2)|^2\,
\sigma_{\rm dipole}(x,{\bm r}),\,\,\,\,\ee
where, $\Psi_{T,L}$ are  the light-cone wavefunctions of $\gamma^*$ 
with transverse, or longitudinal, polarization,  
and $\sigma_{\rm dipole}(x,{\bm r})$ is the cross-section 
for dipole--proton scattering (for a dipole of transverse size ${\bm r}$),
containing all the information about hadronic interactions such as 
the unitarization or saturation effects. 

A simple parametrization for $\sigma_{\rm dipole}(x,{\bm r})$ 
which has qualitatively plausible behaviors like color transparency 
and saturation effects was first proposed by 
Golec-Biernat and W\"usthoff (GBW) \cite{GBW}. 
They used a very simple function 
$\sigma_{\rm dipole}(x,{\bm r})=
\sigma_0\, (1 - {\rm e}^{-{\bm r}^2 Q_s^2(x)/4})$
with only three parameters 
$\sigma_0$ (a hadronic cross-section), $x_0$ and $\lambda$ for 
the saturation momentum $Q_s^2 (x)= (x_0/x)^\lambda$ GeV$^2$, and 
managed to provide rather good fits to the (old) HERA data for 
$x \le  10^{-2}$ and all $Q^2$. This success was quite impressive 
by itself since it suggested the relevance of saturation physics 
in the HERA data, but at the same time required more serious theoretical 
effort towards understanding the HERA data with the saturation picture 
better rooted in QCD. In fact,  there is no kinematical 
regime in which the GBW model can be (strictly) justified from 
QCD, and the GBW model must be improved with the information 
of QCD, or replaced by other QCD-based parametrization.
So far, there are several attempts to improve the GBW model 
\cite{BGBK,KT}, but they mostly focused upon improving 
the behavior of the fit at large $Q^2$, by including DGLAP 
dynamics. On the other hand, we know that the BFKL dynamics, rather than 
DGLAP, should be the right physics in the transition regime towards 
saturation. This BFKL physics was not addressed so far, and we will
focus on the regime with not too large $Q^2$ where the BFKL and 
saturation physics should be more relevant, and will present a 
new analysis of the HERA data, which is rather orthogonal to the 
previous attempts to improve the GBW model.

\section{The CGC fit \cite{IIM}}
\vspace*{-3mm}
We restrict ourselves to the kinematical range where one expects 
important high density effects --- namely, $x\le 10^{-2}$ 
and $Q^2 < 50\,{\rm GeV}^2$ ---,
and show that the data in this range are consistent with our present
understanding of CGC (BFKL evolution and saturation). 
The upper limit 
on $Q^2$ has been chosen large enough to include a significant 
number of ``perturbative'' data points, but low enough to 
justify the emphasis on BFKL, rather than DGLAP, evolution.
Within this kinematical range,  we shall provide a reasonable fit 
(which we call the ``CGC fit'')
to the new HERA data for $F_2$ based on a simple, analytic, 
formula for the dipole scattering amplitude.

\begin{figure}
  \begin{center}
  \includegraphics[width=0.50\textwidth]{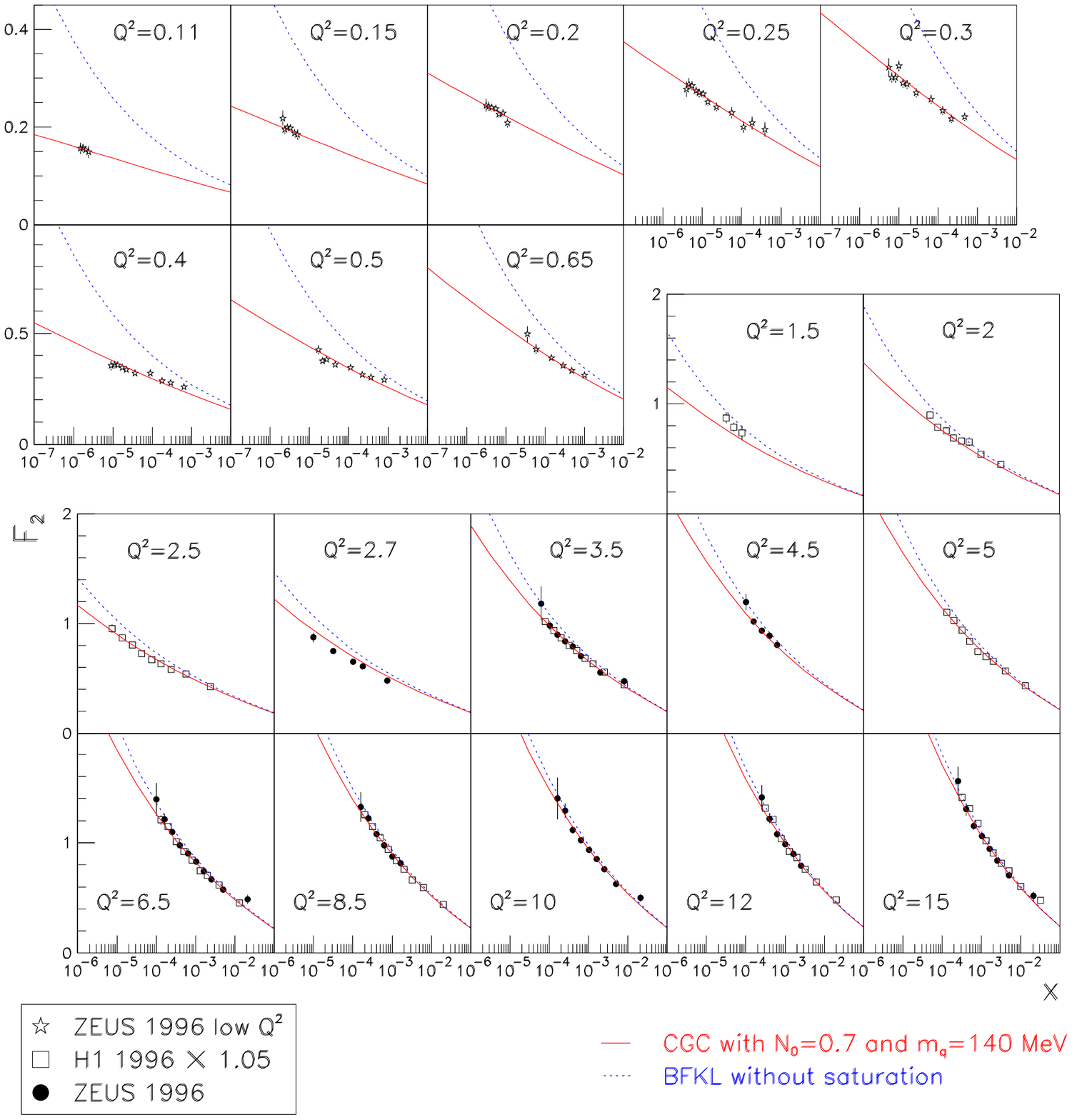}\hspace*{-5mm}
  \includegraphics[width=0.50\textwidth]{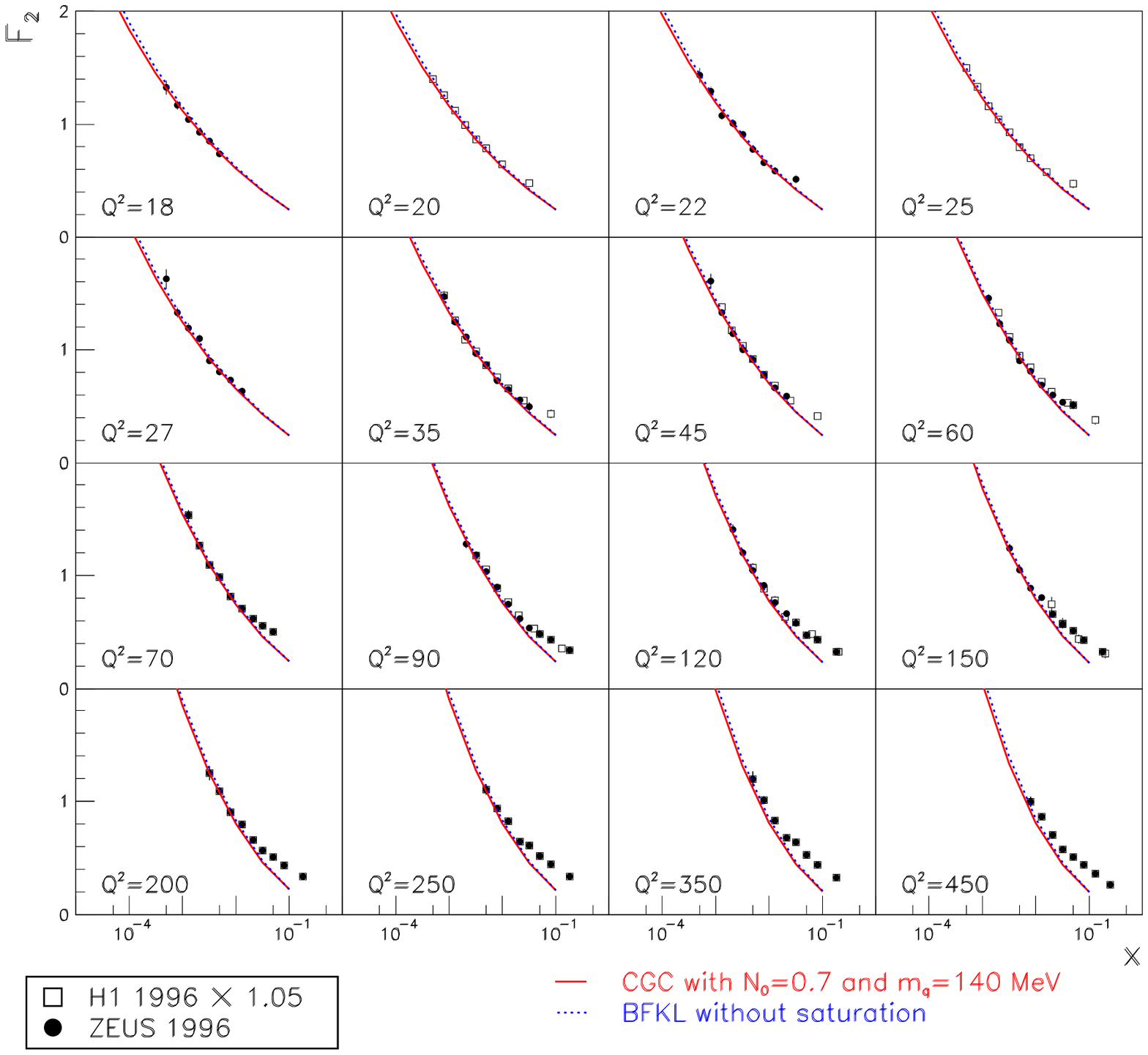}
  \vspace*{-0.7cm}
  \end{center}
 \caption[]{The $F_2$ structure function as a function of $x$ 
in bins of $Q^2$ for $Q^2\le 15\, {\rm GeV}^2$ (left) and 
for $Q^2>15\, {\rm GeV}^2$ (right).
The experimental points are the latest 
data from the H1 and ZEUS collaborations \cite{New_HERA}.
The full line shows the result of the CGC fit with 
${\mathcal N}_0=0.7$ to the ZEUS data for $x\leq 10^{-2}$ and 
$Q^2\leq 45\ \mbox{GeV}^2$. 
The dashed line shows the predictions of the pure BFKL part of the fit 
(no saturation). In the bins with $Q^2\geq 60\, {\rm GeV}^2$, 
the CGC fit is extrapolated outside of the range of the fit.}
\end{figure}

The dipole cross-section in the CGC fit reads 
$\sigma_{\rm dipole}(x,{\bm r})=2\pi R^2 {\mathcal N}(rQ_s,Y)$, where 
the dipole scattering amplitude ${\mathcal N}(rQ_s,Y)$ is constructed 
by smooth interpolation between two limiting solutions to 
the non-linear evolution equations in QCD \cite{BKW,PI}: 
the solution to the BFKL equation with saturation boundary
\cite{EIM02,MT02,DT02} for small dipole sizes, 
$r\ll 1/Q_s(x)$, and the Levin-Tuchin law \cite{LT99,SAT} for 
larger dipoles, $r\gg 1/Q_s(x)$. 
Namely, 
\vspace*{-0.6cm}
\be\label{NFIT}
{\mathcal N}(rQ_s,Y)
&=&{\mathcal N}_0\, \left(\frac{{r} Q_s}
{2}\right)^
{2\big\{\gamma_s + 
\frac{\ln(2/rQ_s)}{\kappa \lambda Y}\big\}}\,\,\quad{\rm for}\quad rQ_s\le 2,
\nonumber\\
{\mathcal N}(rQ_s,Y)&=& 1 - {\rm e}^{-a\ln^2(b\, rQ_s)}\qquad\quad\qquad
{\rm for}\quad rQ_s > 2,
\ee
where $Y=\ln(1/x)$, $Q_s\equiv Q_s(x) = (x_0/x)^{\lambda/2}$ GeV,
and we have defined $Q_s$ in such a way that 
${\mathcal N}(rQ_s,Y)={\mathcal N}_0$ for $rQ_s=2$. 
 The coefficients $a$ and $b$ are 
determined uniquely from the continuity of 
${\mathcal N}(rQ_s,Y)$ at $rQ_s=2$.
In the first line, $\gamma_s=0.63$ (or more strictly, $1-\gamma_s$) 
is the anomalous dimension, and $\kappa=\chi"(\gamma_s)/\chi'(\gamma_s)
\simeq 9.9$ is the diffusion coefficient. The anomalous dimension gives 
the geometric scaling \cite{geometric,EIM02,MT02}, while the second 
"diffusion" term in the power (the term depending upon $\kappa$) brings 
in scaling violations. 
The overall factor ${\mathcal N}_0$ 
is ambiguous, reflecting an ambiguity in the definition of $Q_s$. 
But the results of the fit do not change largely by changing  
${\mathcal N}_0$. 
The saturation exponent $\lambda$ is computable in QCD 
(known up to the renormalization-group-improved NLO BFKL) 
\cite{EIM02,MT02,DT02}, 
but we treat $\lambda$ as a free parameter since the results are
sensitive to its precise value.
We work with three quarks of equal mass $m_q$ and 
use the same photon wavefunctions $\Psi_{T,L}$
as in Refs.~\cite{GBW,BGBK,KT}. 
Thus, the only free parameters of the CGC fit are $R$, $x_0$ and $\lambda$,
which are the same as in the GBW model ($\sigma_0=2\pi R^2$).

The CGC fit has been performed for the $F_2$ data at ZEUS \cite{New_HERA}
with $x\leq 10^{-2}$ and $Q^2$ between 0.045 and 45 $\mbox{GeV}^2$ 
(156 data points). In Fig. 1, the results of the fit are plotted 
against the data for ${\mathcal N}_0=0.7$ and $m_q=140$ MeV. 
The three parameters are determined to be $R=0.641\, $fm, 
$x_0=0.267\times 10^{-4}$ and $\lambda=0.253$ with $\chi^2/$d.o.f.$\, =0.81$.
We also show (with dashed line) the prediction of the BFKL 
calculation without saturation, as obtained by extrapolating the 
formula in the first line of Eq.~(\ref{NFIT}) to arbitrarily 
large $rQ_s$.
This pure BFKL fit shows a too strong increase with $1/x$ at small $Q^2$.
On the other hand, the complete fit, including saturation, works 
remarkably well even at the lowest values of $Q^2$ that we have included.
Note also that the value $\lambda=0.25$ determined from the fit 
is in good agreement with the theoretical result \cite{DT02}. 
We have done the fit separately with the pure scaling part,  
and found that the fit becomes worse. This means that 
the ``diffusion'' term which violates the geometric scaling 
is crucial to fit the HERA data. This is not surprising  because
this term effectively changes the anomalous dimension from its BFKL 
value $\gamma_s=0.63$ (for relatively large dipole sizes $\simle 1/Q_s$) 
to the DGLAP value $\gamma=1$ (for small dipole sizes), and thus 
partially simulates the DGLAP dynamics. However, as is evident from 
the figure, there is a deviation between the CGC fit and the data 
at high $Q^2$ and not so small $x$. This is again not surprising 
simply because this regime is outside the range of validity of the 
CGC fit, which does not include the DGLAP physics (in its right form) 
nor valence quark dynamics.

\section*{Acknowledgments}
\vspace*{-3mm}
The author would like to thank Edmond Iancu and Stephane Munier
for the collaboration whose results are presented in the talk.\\

\end{document}